\documentclass[manuscript]{IEEEtran}

\IEEEoverridecommandlockouts

\IEEEpubid{\makebox[\columnwidth]{979-8-3503-7550-3/24/\$31.00 \copyright 2024 European Union \hfill}\hspace{\columnsep}\makebox[\columnwidth]{}}

\usepackage{cite}
\usepackage{amsmath,amssymb,amsfonts}
\usepackage{algorithmic}
\usepackage{graphicx}
\usepackage{textcomp}
\usepackage{xcolor}
\usepackage{multicol}
\usepackage{textcomp}
\usepackage{wrapfig}
\usepackage{tikz,xcolor,hyperref}
\usepackage{url}
\usepackage{booktabs}
\usepackage{xcolor}
\usepackage{soul}
\usepackage{svg}
\usepackage{pdfpages}

\def\BibTeX{{\rm B\kern-.05em{\sc i\kern-.025em b}\kern-.08em
    T\kern-.1667em\lower.7ex\hbox{E}\kern-.125emX}}
\begin{document}

\title{Examining the Effects of Human-Likeness of Avatars on Emotion Perception and Emotion Elicitation\\
}

\author{
\begin{minipage}[t]{0.3\textwidth}
  \centering
  \IEEEauthorblockN{1\textsuperscript{st} Shiyao Zhang}\\
  \IEEEauthorblockA{\textit{Cognitive Systems Lab} \\
  \textit{University of Bremen}\\
  Bremen, Germany \\
  0000-0001-5965-0428}
\end{minipage}
\hfill
\begin{minipage}[t]{0.3\textwidth}
  \centering
  \IEEEauthorblockN{2\textsuperscript{nd} Omar Faruk}\\
  \IEEEauthorblockA{\textit{Digital Media Lab} \\
  \textit{University of Bremen}\\
  Bremen, Germany \\
  0009-0004-9991-7456}
\end{minipage}
\hfill
\begin{minipage}[t]{0.3\textwidth}
  \centering
  \IEEEauthorblockN{3\textsuperscript{rd} Robert Porzel}\\
  \IEEEauthorblockA{\textit{Digital Media Lab} \\
  \textit{University of Bremen}\\
  Bremen, Germany \\
  0000-0002-7686-2921}
\end{minipage}

\vspace{1em}

\begin{minipage}[t]{0.3\textwidth}
  \centering
  \IEEEauthorblockN{4\textsuperscript{th} Dennis Küster}\\
  \IEEEauthorblockA{\textit{Cognitive Systems Lab} \\
  \textit{University of Bremen}\\
  Bremen, Germany \\
  0000-0001-8992-5648}
\end{minipage}
\hfill
\begin{minipage}[t]{0.3\textwidth}
  \centering
  \IEEEauthorblockN{5\textsuperscript{th} Tanja Schultz}\\
  \IEEEauthorblockA{\textit{Cognitive Systems Lab} \\
  \textit{University of Bremen}\\
  Bremen, Germany \\
  0000-0002-9809-7028}
\end{minipage}
\hfill
\begin{minipage}[t]{0.3\textwidth}
  \centering
  \IEEEauthorblockN{6\textsuperscript{th} Hui Liu*}\\
  \IEEEauthorblockA{\textit{Cognitive Systems Lab} \\
  \textit{University of Bremen}\\
  Bremen, Germany \\
  0000-0002-6850-9570\\
  hui.liu@uni-bremen.de}
\end{minipage}
}
\maketitle

\begin{abstract}
Digital communication has become a fundamental part of our daily interactions and professional pursuits. 
An increasing number of online interaction settings now provide the possibility to visually represent oneself via an animated avatar instead of a video stream. 
Benefits include protecting the communicator's privacy while still providing a means to express their individuality. 
In consequence, there has been a surge in means for avatar-based personalization, ranging from classic human representations to animals, food items, and more.
However, using avatars also has drawbacks. 
Depending on the human-likeness of the avatar and the corresponding disparities between the avatar and the original expresser, avatars may elicit discomfort or even hinder effective nonverbal communication by distorting emotion perception. 
This study examines the relationship between the human-likeness of virtual avatars and emotion perception for Ekman's six "basic emotions," which are anger, disgust, fear, happiness, sadness, and surprise.
Research reveals that avatars with varying degrees of human-likeness have distinct effects on emotion perception. 
High human-likeness avatars, such as human avatars, tend to elicit more negative emotional responses from users, a phenomenon that is consistent with the concept of Uncanny Valley in aesthetics, which suggests that closely resembling humans can provoke negative emotional responses. 
Conversely, a raccoon avatar and a shark avatar, known as cuteness, which exhibit moderate human similarity in this study, demonstrate a positive influence on emotion perception.
Our initial results suggest that the human-likeness of avatars is an important factor for emotion perception. 
The results from the follow-up study further suggest that the cuteness of avatars and their natural facial status may also play a significant role in emotion perception and elicitation. 
We discuss practical implications for strategically conveying specific human behavioral messages through avatars in multiple applications, such as business and counseling. 


\end{abstract}


\section{Introduction}
\label{sec:introduction}
Digital communication has become a fundamental part of our daily lives and professional pursuits. 
An increasing number of online interaction settings now provide the possibility to visually represent oneself via an animated avatar instead of a video stream. 
The concept of avatars, which function as visual proxies for us across various virtual settings, is central to our digital identity \cite{Lin_Latoschik_2022}. 
The benefits of avatars include protecting the communicator's privacy while still providing a means to express their individuality. 
In consequence, there has been a surge in means for avatar-based personalization, covering a wide variety ranging from simple 2D avatars and classic human representations to animals, food items, and more \cite{kostic_nonverbal_2015}. 
At one end of the scale, we have animated 3D avatars that are uncannily accurate replicas of our actual appearance, using cutting-edge graphics and detailed design features. 
On the other end of the spectrum are simple static 2D avatars, which are reduced to fundamental shapes and forms while bearing great symbolic importance \cite{kostic_nonverbal_2015}. 
Using avatars as a medium of human communication raises the question of whether human interlocutors can perceive the communicator's intentions and emotions as accurately when interacting with a realistically drawn human avatar as when talking to an abstract banana avatar.

Nonverbal human communication relies heavily on facial expression recognition \cite{kappas_facial_2013}. 
Subtle adjustments in face muscles or tiny changes in posture can indicate a lot in physical environments \cite{Ko_2018}. 
Subtle variations like these may be amplified, minimized, or even omitted across various facial features of the given avatars, potentially leading to discrepancies in the accuracy of the conveyed expression. 
The ability of an avatar to effectively transmit emotions, intentions, or moods can have a tremendous impact on user engagement, collaboration, and even the emotional resonance elicited by the digital space \cite{Koles_Nagy_2016}.

However, the interaction between the degree of avatar abstraction and emotion recognition performance remains comparatively unexplored. 
For example, a highly realistic avatar may provide more facial signals, but an abstract avatar may use symbolic or exaggerated features to convey emotion \cite{Ekdahl_Osler_2023}. 
Thus, while a more realistic and human-like appearance may be advantageous in certain socially intense situations, other settings may benefit from more simplified user representations \cite{kostic_nonverbal_2015}.
Different degrees of realism may further modulate the intensity of perceptions of the same message, up to the point that the emotions evoked in the receiver may also change qualitatively.
The small-scale study presented below therefore examines how varying degrees of human-likeness in avatars influence both the perception and elicitation of emotions in users during an interactive virtual environment. 
Moreover, it's imperative to ascertain whether users perceive these emotions conveyed by avatars as a universally understood interpretation or whether the perception varies significantly among different users. This variability, known as inter-observer reliability, is pivotal in assessing the effectiveness of avatars as communication tools.
By understanding the varied emotional perceptions and elicitations evoked by different types of avatars, users can systematically choose them to match specific communication goals in various usage scenarios, thereby the effectiveness of conveying specific human behavioral messages might be enhanced.

\section{\uppercase{Related Work}}
\label{sec:related work}
Avatars that were more anthropomorphic were perceived to be more attractive and credible, and people were more likely to choose to be represented by them \cite{Nowak_Rauh_2005}. 
The presence of an anthropomorphic appearance triggers people’s simplistic social scripts (e.g., politeness, reciprocity), which in turn induce cognitive, affective, and social responses during interactions with technology \cite{Wang_Baker_Wagner_Wakefield_2007}. 
Dynamic virtual avatars has been suggested for studying emotion recognition in a face in that they recreate realistic stimuli in emotion research \cite{10.1007/978-3-319-22888-4_30}.
The level of abstraction in avatar design can affect the recognition of facial expressions. 
Since voice and facial emotions are significant means of expressing emotion \cite{Banse_Scherer_1996},. 
Key emotional cues are communicated through posture and physical movements \cite{Dael_Mortillaro_Scherer_2012}. 
An essential component of nonverbal communication that expresses emotion is gesture \cite{Dael_Goudbeek_Scherer_2013}. 
People can recognize emotions more accurately by integrating gestures \cite{gunes_bi-modal_2007} and facial expressions because they can infer one another's feelings through systems that interpret implicit information \cite{cowie_emotion_2001}. 
Even with minimal information about body movements, emotions can be detected, as in the "point light display" phenomenon \cite{givens2014nonverbal}.

A study on affect-sensitive human-computer interaction analyzed a user's interaction with four different virtual avatars, each manifesting distinct emotional displays, based on the principles of Affect Control Theory \cite{Shang_Zhengkun}. 
The results demonstrated that the probabilistic framework used in the study enabled the system to perceive the user’s and agent’s feelings. 
Another study found that realistic avatars were rated significantly more positively and more human-like than abstract wooden mannequin avatars \cite{Hepperle_Purps_Deuchler_Wölfel_2021}. 
This suggests that the level of realism in avatar design can impact how users can recognize and interpret facial expressions. 
Furthermore, researchers have also discovered that engaging in conversations with virtual humans can serve as a valuable source of support during times of distress \cite{PAUW2022107368}. 
This finding further reinforces the relatively positive impact of realistic avatars. 
The researchers discovered through brain activity analysis that perceiving emotional expressions from human-like characters closely resembles perceiving human emotions, triggering similar behavioral and physiological responses \cite{Borst_de_Gelder_2015}. 
However, when artificial faces closely resemble human faces, the expressed emotion may provoke a reduced response from the observer, indicated by lower intensity scores and decreased brain activity. 
The study demonstrates that brain activity undergoes changes during the perception of emotions in a human-like manner.
One study investigated the role of photographic and behavioral realism of avatars with animated facial expressions on perceived realism and congruence ratings \cite{Grewe_Liu_Kahl_Hildebrandt_Zachow_2021}. 
The study found that avatars with learned facial expressions were rated as more photographically and behaviorally realistic and had a lower mismatch between the two dimensions. 
Another study examined the self-representation of physical, demographic, and personality characteristics through avatars, differences in self-representation between various online activity contexts, and between-participant variance in ascribed personality \cite{Zimmermann_Wehler_Kaspar_2022}. 
The study found a high level of congruence between the demographic and physical characteristics of the avatar, the actual self, and the ideal self. Furthermore, it found an idealized representation of the avatar’s personality traits, which was affected by the specific activity context. 

The influence of human-likeness in the design of the avatar, such as human-like facial expressions and self-presentation characteristics, etc., is evident in their impact on human behavior.
However, further research is needed to fully understand the relationship between avatar design and both emotion perception and emotion elicitation. 
Given that basic emotions can amalgamate to generate complex or compound emotions \cite{ekman1992there}, our study sought to initially investigate the relationship between the perception and elicitation of the six basic emotions and the degree of human-likeness exhibited by avatars.

\section{Methodology}
\label{tab: methodology}

To understand the relationship between avatar design and emotion perception and emotion elicitation, we conducted two studies: a preliminary study with 4 participants viewing 18 emotion videos displayed by 3 avatars and a follow-up study with 11 participants viewing 30 emotion videos displayed by 5 avatars. 
In our annotation-based approach, we used videos of real people acting out six basic emotions. These recordings were then automatically converted and displayed by avatars.
The video recording section discusses the software used in this study and improvements made to the video recording process, which includes manual recording, photo recording, and video recording methods, along with their respective advantages and disadvantages.
In a within-subjects study, all participants conducted the first emotion perception task, followed by the emotion elicitation task. 
The following sections will introduce more details about our study design, emotion video recording, such as content, quantity, etc., and the procedure of data collection.

\subsection{Experiment design}
\label{sec: experiment design}

In the experiment, participant reactions to Ekman's six basic emotions \cite{ekman1999basic} from different avatars were collected. 
Prioritizing the questionnaire design was crucial to ensuring comprehensive, authentic, and unbiased results. 
The questionnaire aimed to collect subjects' opinions regarding emotion videos performed by avatars and was structured around two main tasks: watching emotion videos performed by different avatars and answering two questions for each.
The emotion perception task required participants to respond to the following questions for each video:
\begin{itemize}
    \item What emotion do you believe the avatar displays in the video?
    \item How confident are you in your judgment of the avatar's emotion?
\end{itemize}
In the emotion perception task, participants are required to identify the emotion displayed by the avatar in the video and provide a confidence value indicating their certainty regarding their answer.

In the emotion elicitation task, participants were prompted to address the following two questions for each video:
\begin{itemize}
    \item What emotion do you believe the avatar in the video made you feel?
    \item How confident are you in your judgment of the emotion you felt?
\end{itemize}
In contrast to the emotion perception task, where subjects identified the avatar's expression in the video, the emotion elicitation task required participants to identify the emotion they felt and give a confidence value for the corresponding answer. 
Distinct from the emotion perception task, the purpose of the emotion elicitation task is to explore the effect of different avatars on participants' emotion elicitation. 
In addition to the main task, participants reported their current emotions at the beginning and end of the questionnaire.
This additional measure aimed to provide insights into which emotions may be elicited by different types of avatars.

\subsection{Video recording}
\label{sec: video recording}
To ensure the consistency of emotion expression in videos, the real people acting out the six basic emotion videos are employed in the video recording of our study, which are then automatically transposed into avatars and consequently performed by them. 
Since the conversion remains within the same video format (video-to-video), the emotional cues are not affected by changes in dimensions. 
The following section will detail the software and techniques used in this video recording process.

\subsubsection{Software}
\label{sec: software}
In this study, we used the Avatar software provided by Animaze by Facerig\footnote{https://www.animaze.us/}. 
One of the preliminary goals of Animaze is to facilitate genuine human emotional connection through avatars \cite{The_Animaze_Team_2021}. 
Animaze can use one or more types of trackers to track various aspects of the user's expression and speech. 
To eliminate the influence of speech on emotion perception and elicitation, only expression tracking is utilized in the recording. 
Our study focuses solely on nonverbal communication between the avatar and the user.
Animaze has been not only used in various aspects of research, such as real-time facial motion capture \cite{gomez2023daddy}, analysis of emotion expression impacted by facial features \cite{van2022investigating} and the effects of varying lecturer avatars \cite{mizuho2023virtual}, also employed by some streamers and VTubers during live streams and videos, especially as more and more users during pandemics started live streaming using avatars to obtain more contacts with others while protecting personal privacy. 
It offers a variety of avatars to meet different usage scenarios and users can also personalize their avatar characters through the software's customizable settings. 
Animaze is compatible with webcams or cell phones and functions seamlessly with major video conferencing and social networking sites. 

To represent a range of human-likeness from high to low, we chose avatars depicting a human, raccoon, and banana in a preliminary study. 
These avatars were selected from broader categories that included additional options. Further details about the human-likeness score are detailed in Table \ref{tab:Human-likeness-init}.
Additionally, we aimed to select avatars with broad usability, as those available in Animaze are commonly used across social media platforms like YouTube and TikTok. 
The avatars provided by Animaze are categorized into four groups: human, animal, food, and object avatars.
The human avatar is designed to encompass a broad spectrum of human faces and bodies, ensuring it is readily recognizable as human. 
Animal avatars, on the other hand, span a diverse range of species, each with distinct facial and body features. For instance, the raccoon avatar utilized in this experiment prominently features its characteristic hairy back, with only a few human-like body features visible, thus closely resembling the physical appearance of a real raccoon. Despite an abundance of facial features, the presence of hair complicates the recognition of the raccoon's eyebrows. 
Consequently, the raccoon avatar bears less facial resemblance to humans compared to the human avatar. 
Furthermore, animal avatars vary in their degree of similarity to human physical features; for example, pigs exhibit more human-like characteristics, while sharks diverge significantly. 
These differences will be further explored in subsequent experiments.
Food and object avatars are relatively abstract, often lacking complete facial features and physical attributes. 
Since our study focuses on emotions, facial features play a crucial role in conveying emotions, surpassing the significance of physical features in this context.

\subsubsection{Approach}
We selected avatars capable of clearly expressing the six basic emotions to ensure consistency in the performance of each emotion. 
Our goal was to create consistent source materials for recording the same emotions across human, raccoon, and banana avatars, thereby minimizing any potential impact of video quality on emotion recognition results.
The following methods were all attempted in the experiment:
\begin{itemize}
    \item Source material involves a person displaying an expression in front of the camera.
    \item Recording from photos depicting the six basic emotions.
    \item Recording from pre-existing videos featuring the six basic emotions.
\end{itemize}

\begin{figure}[htbp]
\centerline{\includegraphics[width=\linewidth]{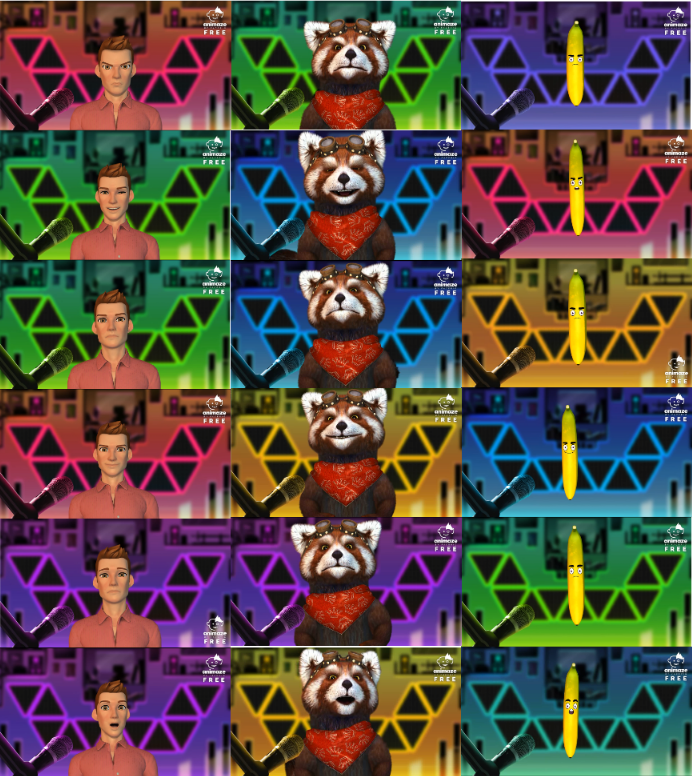}}
\caption{Screenshots of all the avatar videos from the preliminary experiments, which included three avatars (from left to right: human, raccoon, banana). Each avatar performed six basic emotions (from top to bottom: anger, disgust, fear, happiness, sadness, and surprise).}
\label{fig:all videos}
\end{figure}

Manual recording offers precise control over expression, which is crucial with non-actors. 
Precise manipulation of the five senses to convey emotions accurately and maintain consistency in each repetition of the expression presents significant challenges. 
Furthermore, subtle, involuntary movements such as uncontrollable blinking and gaze shifts, which are natural in human communication, can be difficult to replicate perfectly and might hinder the effectiveness of the recording. 
One of the obvious limitations of using the software is that eye movement recognition can be inaccurate for users with glasses or smaller eyes, resulting in the avatar constantly blinking or squinting. 
Since eyes play a vital role in emotional expression, this imprecision may evolve into a barrier to emotion perception.

Taking into account the potential impact of subtle changes in videos on emotion recognition, an attempt was made to utilize photographs designated to represent various emotional states as source material. 
However, due to the inherent 2D nature of photographs, the process of deriving emotions often entails a direct transition from the natural expression to a specific frame in the emotional expression sequence. 
These abrupt transitions result in emotion videos that appear raw and deviate from the natural evolution of the expression.  
Consequently, this limitation leads to recorded videos that fail to adequately convey the nuances of facial expressions.

Instead of working with generic 2D data, our approach therefore used pre-defined videos of the six basic emotions recorded by real people as the source material for Avatar video recording.
Sourced from the YouTube channel "Science of People", these videos ensured an authentic portrayal of the corresponding emotions. 
Each expression in these videos lasted only 1-2 seconds, providing ample time for subjects to recognize the emotions by repeatedly viewing a single video where each expression was repeated several times (6-8) in the avatar video \footnote{https://forms.gle/h88TBg1Ui3J2bKrc7}. 
We set the video to repeat and stabilize it to the front of the computer camera. 
Each recording was kept at a uniform angle and distance. 
The variations in facial expressions recognized by the avatars were obtained from pre-recorded videos, which had been annotated with different emotions by the publisher. 
The videos show expressions of two stationary performers whose facial features are well recognized without the blinking and squinting problems described above.
To standardize the questionnaire, the length of each video was set between 8 and 12 seconds, allowing participants to replay the videos until they identified the answers to the relevant questions as they completed the questionnaire.

\subsection{Data collection}
\label{sec: data collection}
To facilitate the entire experiment online, all videos were embedded directly into the questionnaire, which participants accessed through a distributed link. All responses were collected anonymously. 
5 complete datasets were gathered from 3 male and 2 female participants. 
Each dataset included 36 judgments with confidence ratings, resulting in 180 data points. 
One dataset was excluded from further analysis as it was received after the data analysis was complete. 
The experiment was hosted on Google Forms. 
To ensure the validity of the user study results, participants were contacted after the trial and confirmed to have watched all 36 embedded videos in full, and no technical issues arose during the experiment. 
As a result, all opinions expressed by the four experimental raters included in the statistical analysis are considered valid. 
Further details regarding the data analysis are provided and discussed in Section \ref{sec: results}.

\section{Results of preliminary study}
\label{sec: results}
This section presents the analysis of our small-scale annotated data, highlighting key findings that will guide further research. 
The annotation results were provided by four raters, with each rater annotating all videos once, fulfilling the prerequisite for calculating Fleiss' kappa.
Firstly, the scientific analysis results of the human-likeness of avatars, as predicted by the estimator \url{http://www.abotdatabase.info/predictor}, are discussed in Section \ref{sec: human-likeness}. 
The human-likeness score of the avatar plays a crucial role in the subsequent discussion.
Subsequently, agreements among annotated results for different avatars are analyzed, along with the impact of avatars on emotion perception results, considering the self-confidence value provided by the raters during the annotation process.
Furthermore, an in-depth analysis is conducted to determine whether there are variations in the influence of avatars on the perception of the six basic emotions, as well as their potential effect on eliciting emotions in the participants.
Lastly, the six emotions are categorized into positive and negative categories, examining whether avatars evoke positive or negative emotions in the participants.

\subsection{Human-likeness evaluation of avatar}
\label{sec: human-likeness}
An anthropomorphic analysis was initially conducted on three different avatars, with the results presented in Table \ref{tab:Human-likeness-init}. 
Anthropomorphism refers to “the extent to which an image looks human” \cite{Nowak_Rauh_2005}. 
A tool of robots’ human-likeness is used in this study \textit{the ABOT (Anthropomorphic roBOT) Database} \footnote{https://www.abotdatabase.info/ (accessed April 29, 2024)}. 
ABOT provides researchers and designers with images of 200 real-world anthropomorphic robots built for research or commercial purposes, and it accumulates data on people’s perceptions of this wide variety of robots \cite{9473560}. 
Table \ref{tab:Human-likeness-init} shows the results of human-likeness generated by ABOT. 
The human-likeness estimation scores are as follows: human, raccoon, and banana, arranged in descending order.
Regarding facial features, humans and raccoons share the same facial feature dimension, while the banana has the lowest facial feature dimension due to its inability to distinguish the head from the trunk. 
The ranking of the surface-look dimension aligns with the human-likeness estimation scores, as raccoons and bananas have fur and peels covering their surfaces instead of skin characteristics. 
Additionally, compared to raccoons, bananas lack noses and ears, indicating that bananas exhibit fewer human characteristics, followed by raccoons.
This corresponds to their level of abstraction: the banana avatar possesses the highest degree of abstraction, while the raccoon avatar is more abstract than the human avatar. 
This implies that the human-likeness analysis of the avatar is inversely proportional to its level of abstraction.
Since human-likeness is related to how abstract an avatar is, we'll focus on analyzing human-likeness ratings in the following sections. 
This decision is based on the understanding that the relationship between abstraction degree and emotion perception can be inferred from the relevant results of the human-likeness of avatars and emotion perception.
In the results analysis section, the influence of personification on emotion perception will be further discussed.

\begin{table}[htbp]
    \caption{Human-likeness Estimator Results of three Avatars}
    \label{tab:Human-likeness-init}
    \centering
    \begin{tabular}{@{\extracolsep\fill}rcccc}
        \toprule
        & \multicolumn{3}{@{}c@{}}{Estimator Results} \\\cmidrule{2-4}
        & Human & Raccoon & Banana \\
        \midrule
        Human-likeness score & 74.56 & 48.07 & 9.03 \\
        Body-Manipulators & 0.4 & 0.2 & 0.2\\
        Surface-Look  & 1 & 0.43 & 0.14 \\
        Facial Features  & 1 & 1 & 0.75 \\
        \bottomrule
    \end{tabular}
\end{table}

Human-likeness score interpretation:
\begin{itemize}
    \item The overall human-likeness score ranges from 0 (not human-like at all) to 100 (just like a human).
    \item The dimension scores are the average ratings of all marker items in each dimension:
    \begin{itemize}
        \item Body-Manipulators: mean (torso, legs, arms, hands, fingers)
        \item Surface Look: mean (gender, head hair, skin, nose, eyebrow, eyelashes, apparel)
        \item Facial Features: mean (head, face, eyes, mouth)
    \end{itemize} 
\end{itemize}

\subsection{Reliability of agreements of emotion perception}
\label{sec: reliability}
The statistical method of Fleiss' kappa is used to evaluate the reliability of consistency among a fixed number of raters when assigning classification ratings to multiple projects or classifying projects. 
This method is applied to binary or nominal scales. 
A high Fleiss' kappa indicates strong agreement among raters, while a low Fleiss' kappa suggests rater disagreement. 
We calculated the Fleiss' Kappa values for various avatars expressing the six basic emotions. 
This helped us investigate the correlation between the reliability of participants' agreement with different avatars and emotions and whether the agreement was related to the avatar's human-likeness.




In the emotion perception task, we observed fair agreement for all emotions and avatars. Participants in this study showed substantially more variability, with all Fleiss' kappa values below 0.4.
The Fleiss' kappa value suggests that participants have substantial agreement with the raccoon avatar with a Fleiss’ kappa of 0.81, while it was observed in the results that the confidence value of raccoons' emotion perception is higher.
Since the preliminary study provided only ideas for analysis and rough relationships between avatar's human-likeness and emotion perception and elicitation, more detailed results supported by statistical tests are introduced in follow-up-extended experiments in Section \ref{sec: Follow-up extended study}.
The results of emotion perception appear to suggest that the participants have higher confidence in the perception consistency of the raccoon avatar's emotional communication. 
The consistency and reliability are medium for human avatars, which is half that of raccoon avatars.
According to the human-likeness score of the banana, the banana avatar is the embodiment with the highest level of abstraction among the three avatars; its human-likeness score is the lowest, as shown in Section \ref{sec: human-likeness}. The low human-likeness value might bring more ambiguity to emotion perception. 
Additionally, by comparing the results between raccoons and human avatars, it can be concluded that the ambiguity in emotion recognition may not decrease with the increase of human-likeness value. 

Comparing the confidence values of the two tasks (emotion perception and emotion elicitation) based on observation, the results suggest that raters are more confident in judging the emotions the avatars portrayed (Task 1: Emotion perception) than the emotions they felt (Task 2: Emotion elicitation). 
In the annotation results for the banana avatar, raters exhibit lower confidence levels in both recognizing the avatar's emotions and their own emotional responses. 
This indicates that the banana avatar has caused more confusion among raters in emotion perception. 
Conversely, the results from the raccoon avatar suggest that its emotional communication is more accurate and consistent in its emotional influence. 
Additionally, in the task of emotion elicitation, human avatars elicited more confident emotional responses in raters compared to banana avatars.

\subsection{Ambiguity among emotions}
\label{sec: ambiguity}
In the emotion elicitation task, we observed substantial agreement among the 4 participants for expressions of disgust, with a Fleiss’ kappa of 0.778, suggesting a surprisingly high degree of consensus. 
For all other emotions, participants in this study showed substantially more variability, with all other Fleiss' kappa values below 0.4.



\begin{figure}[htbp]
\centerline{\includegraphics[width=\linewidth]{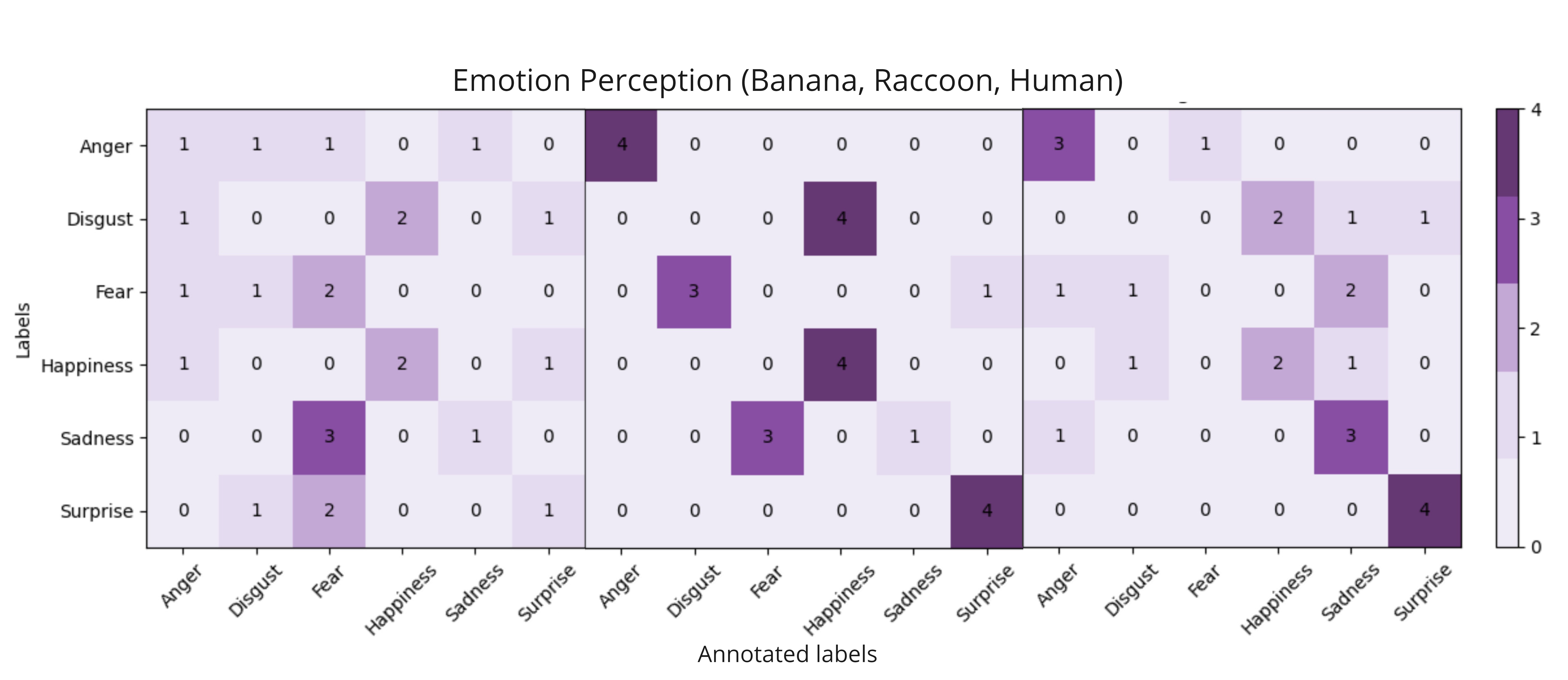}}
\caption{Confusion matrix: Emotion perception among 3 avatars (left to right: banana, raccoon, human) with six basic emotions (left to right and top to bottom: anger, disgust, fear, happiness, sadness, surprise).}
\label{tab: cm_emotion_perception_3}
\end{figure}

Combined with the results of the emotion perception confusion matrix with 6 emotions of three avatars in Figure \ref{tab: cm_emotion_perception_3}, it can be observed that the surprise emotions conveyed by raccoons and humans are perfectly recognized by participants, but for the surprise emotions conveyed by banana avatars, the annotations by raters are not consistent and easily confused with fear. 
Similarly, the emotion with the highest consistent emotion perception is disgust, and raters also have higher confidence in their annotation. 
However, by analyzing the results of the emotion perception confusion matrix with 6 emotions of the three avatars, we found that the rater did not correctly recognize the disgusting emotions that the video wanted to convey, but more consistently recognized the disgust as happiness. 
This might be caused by that the expression of disgusting emotions is not accurately performed by avatars.

\begin{figure}[htbp]
\centerline{\includegraphics[width=\linewidth]{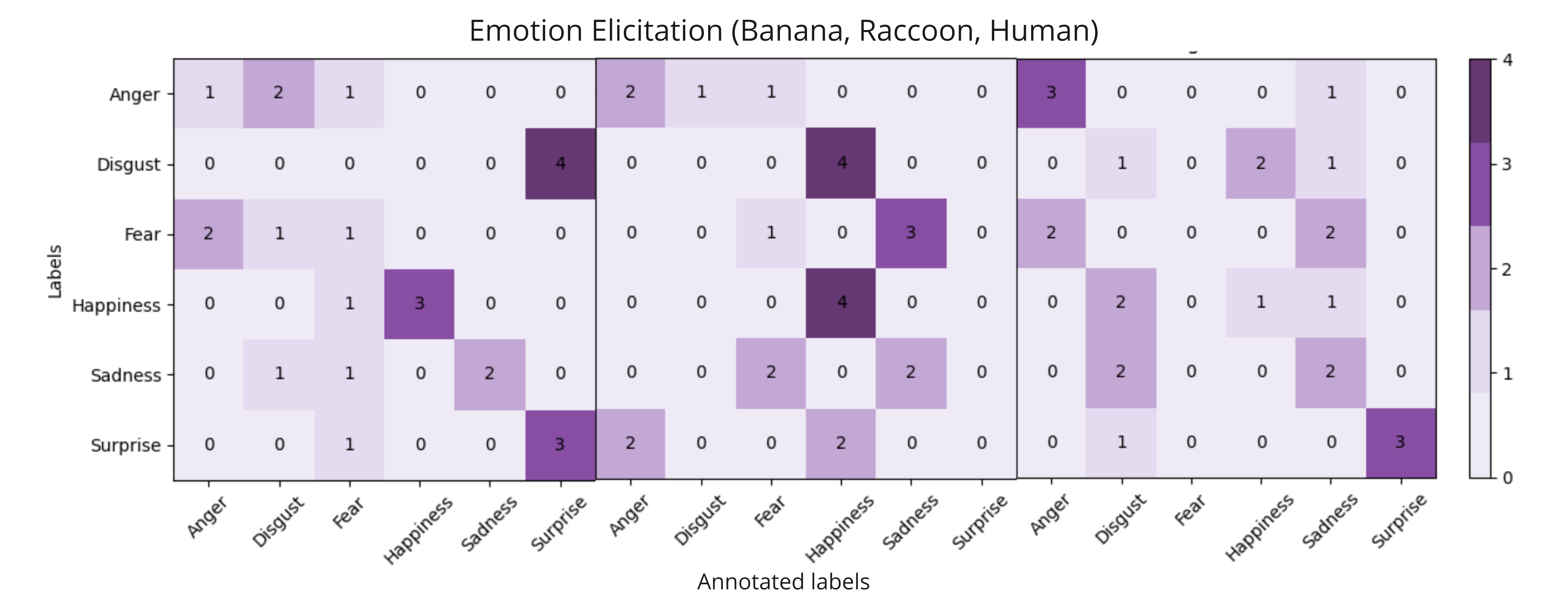}}
\caption{Confusion matrix: Emotion elicitation among 3 avatars (left to right: banana, raccoon, human) with six basic emotions (left to right and top to bottom: anger, disgust, fear, happiness, sadness, surprise).}
\label{tab: cm_elicitation_3}
\end{figure}

In the task of emotion elicitation, the annotation results in Figure \ref{tab: cm_elicitation_3} suggest different characteristics. 
When the avatar conveys disgusting emotions in the video, raters have substantial agreement with the recognition results of their own emotions and are confident in the perception results they have made. 
The results of the confusion matrix show that this video mainly brings happiness and surprise to the rater's own emotions, in which raccoon avatars and human avatars are more likely to make the rater feel happy, while banana avatars can surprise the raters. 
"Disgust" brought a positive emotional influence on raters through the expression of avatars.

"Surprise" gives the rater a high degree of consistent emotion influence, which is similar to the results obtained in emotion perception. 
Surprise emotions tend to bring more consistent recognition results, whether in emotion perception or emotion elicitation. 
However, after watching the video that conveyed the surprise, the raters did not have much confidence in the results of emotion perception. 
The surprise emotions expressed by human avatars and banana avatars can also surprise the rater. 
The surprised emotion expressed by the raccoon evoked happiness and anger in the raters. 
Although the accuracy of the raccoon's emotion perception is the highest, there are still great differences in the emotional response of the raccoon avatar.

\subsection{Emotion induction between positive and negative emotion}
Drawing from the conclusion of the preceding section, a distinction emerges between the accuracy of emotion perception and elicitation by raters. 
While surprise and anger can be identified with precision, the remaining four emotions often lead to confusion. Among the six basic emotions, happiness is categorized as a positive emotion, while the remaining five emotions are classified as negative based on the literature \cite{Ekman_1972}. 

The results reveal that raccoon avatars tend to shift perception from negative to positive emotions. 
Conversely, human avatars prompt a shift in perception from positive to negative emotions during avatar emotion assessment. 
Comparatively, raters find it challenging to perceive the positive emotions conveyed by human avatars, as opposed to raccoon avatars. 
Additionally, in the process of emotion elicitation, raters report experiencing negative emotions during the transition from the banana avatar to positive emotions, suggesting that the banana avatar may induce a negative emotional response.







\section{Follow-up extended study}
\label{sec: Follow-up extended study}
In Section \ref{sec: results}, the results of the analysis of the small-scale preliminary study were summarized. 
In order to validate the generalizability of the findings mentioned in the results chapter and to address the limitations of the study (e.g., four participants can only illustrate the specificity of the findings, but are not generalizable; the variation in human similarity as depicted by three different avatars is not comprehensive, etc.), the idea of a follow-up extended study was proposed. 
Five avatars were used instead of three, and the opinions of 11 new participants were collected. 
Out of these 11 participants, four are female and seven male (\textit{Mean} age = 26.64; \textit{SD} = 3.497), with six participants from Germany, and the remaining five are from China, Iran, India, and Bangladesh. Four of them had previously encountered the avatars mentioned in the text on other social media platforms, none of whom were from Germany. 
The remaining seven had never seen these avatars before participating in the experiment. 
All participants indicated that they had never before interacted with any avatars like the ones used in the present work. 
However, most of the participants from Asian reported having seen the shark on social media. 
In this section, the differences between the extended and preliminary experiments will be detailed, and the results collected from the two experiments will be compared.

\subsection{Differences in study design}
\begin{figure}[htbp]
\centerline{\includegraphics[width=\linewidth]{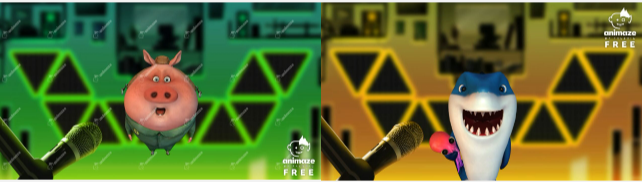}}
\caption{New avatars added only in the follow-up study: Pig and sharks' screenshots from their avatar videos.}
\label{tab: new_avatar}
\end{figure}

Firstly, in contrast to the preliminary experiment, which utilized three avatars, a follow-up study introduced two additional avatars in Figure \ref{tab: new_avatar} to see if the trends found in our preliminary studies persisted or diminished. 
Their human similarity indices are illustrated in Table \ref{tab:Human-likeness}. 
The human similarity index of the pig falls between that of the human and the raccoon, while its facial feature dimension equals that of the human and the raccoon. 
Similarly, the shark's human similarity index lies between that of the raccoon and the banana, with its facial feature dimension matching those of the raccoon, the pig, and the human. 
Human similarity gradually decreases from the human to the banana, with only the banana exhibiting lower facial features compared to the other avatars.
Animal avatars encompass creatures ranging from hairy, such as raccoons and dogs, to hairless, such as sharks. 
Sharks and raccoons differ notably in their human likeness scores, with sharks possessing physical features distinct from humans, while raccoons exhibit more human-like attributes such as clearly defined limbs and clothing. 
The pig avatar was chosen for its complete body representation compared to the raccoon, with body parts resembling human physical characteristics.

\begin{table}[htbp]
    \caption{Human-likeness Estimator Results of five Avatars}
    \label{tab:Human-likeness}
    \centering
    \begin{tabular}{@{\extracolsep\fill}rcccccc}
        \toprule
        & \multicolumn{5}{@{}c@{}}{Estimator Results} \\\cmidrule{2-6}
        & Human & Pig & Raccoon & Shark & Banana \\
        \midrule
        Human-likeness score & 74.56 & 65.96 & 48.07 & 37.26 & 9.03 \\
        Body-Manipulators & 0.4 & 1 & 0.2 & 0.4 & 0.2\\
        Surface-Look  & 1 & 0.43 & 0.43 & 0.14 & 0.14 \\
        Facial Features  & 1 & 1 & 1 & 1 & 0.75 \\
        \bottomrule
    \end{tabular}
\end{table}

Secondly, due to some ambiguity in the way the second task was formulated in the initial experiment, the prompt for emotion elicitation was reformulated in the follow-up study. The formulation was as follows:
\begin{itemize}
    \item Which emotion do you feel in response to the avatar?
    \item How confident are you in your judgment of the emotion you felt?
\end{itemize}
The purpose of emotion elicitation was to explore the effects of the avatar's emotions on the user's perception of emotions, potentially motivating the user's other emotions.
Additionally, several questions about users’ choice of avatars and their reasons were included at the end of the questionnaire. 
These questions were designed to explore user behavior regarding avatar choice. For instance:
\begin{itemize}
    \item Which avatar do you think is the funniest?
    \item Which avatar would you pick for yourself in a casual setting?
    \item Do you prefer to use a more human-like avatar than an abstract avatar?
\end{itemize}

\subsection{Results}
\label{sec: Results}
In the following sections, we present the findings of a subsequent study, which involved 11 participants and incorporated additional statistical analyses, focusing on two aspects: emotion perception and emotion elicitation.
\begin{figure*}[htb]
\centerline{\includegraphics[width=1\textwidth]{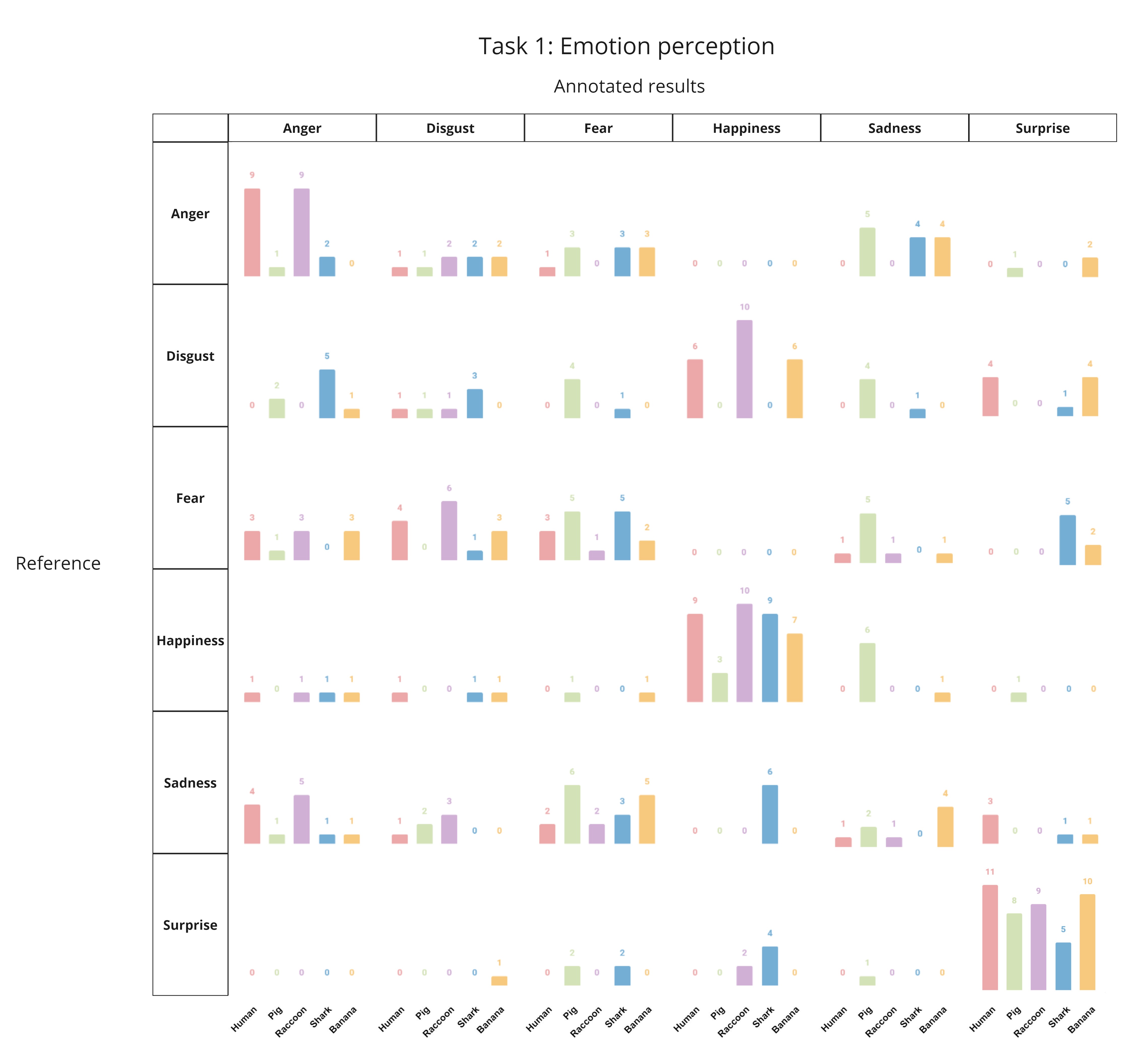}}
\caption{Results of emotion perception rated by 11 participants.}
\label{tab: Emotion perception}
\end{figure*}

\begin{figure*}[htb]
\centerline{\includegraphics[width=1\textwidth]{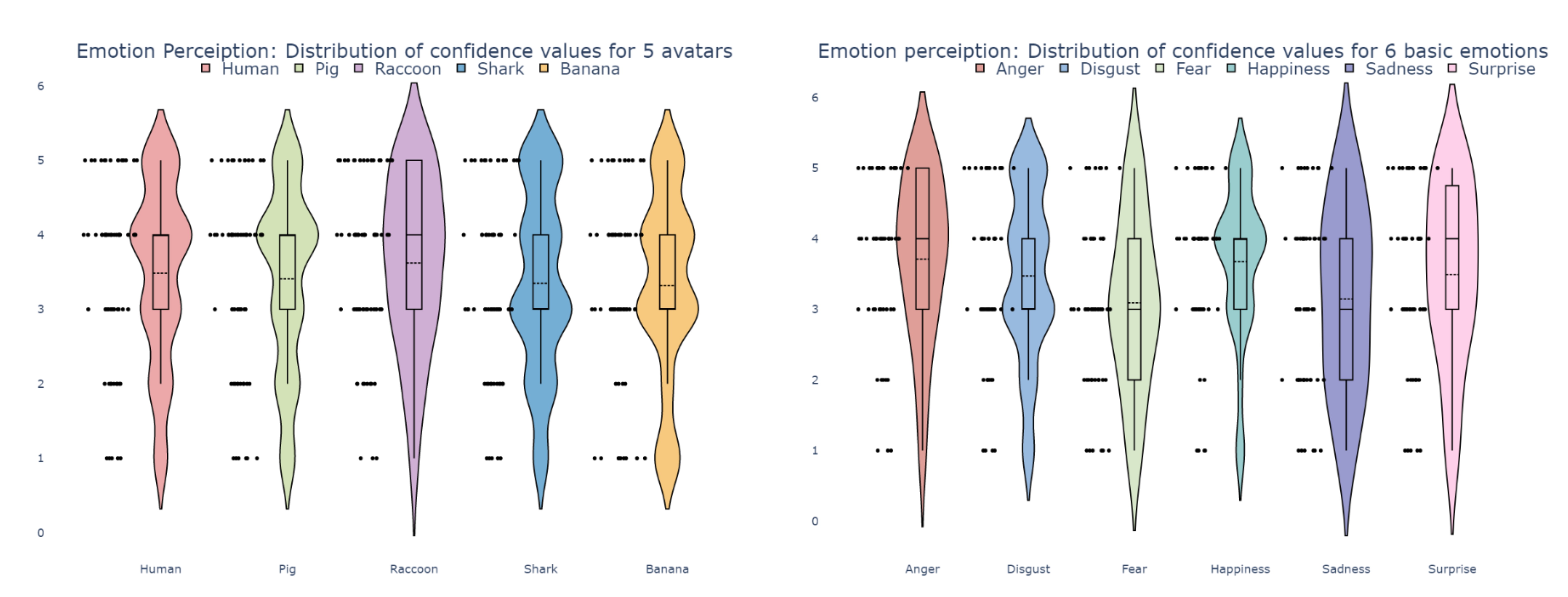}}
\caption{The results of confidence values from 11 participants in task "Emotion perception" are shown in violin plots. Left: The results are analyzed based on 5 different avatars. From left to right: human: red, pig: green, raccoon: purple, shark: blue, banana: yellow. Right: The results are analyzed based on 6 basic emotions. From left to right: anger: red, disgust: blue, fear: light green, happiness: green, purple: sadness, pink: surprise.}
\label{tab: Violin plots emotion perceiption}
\end{figure*}

\subsubsection{Emotion perception}
Figure \ref{tab: Emotion perception} details the distribution of subjective ratings in the emotion perception task.
A repeated measures analysis of variance was performed with emotion (fear, disgust, happiness, anger, surprise, sadness) and avatar (banana, bear, human, pig, shark) as independent variables on the two tasks as dependent measures (perception, elicitation). 
While the main effect of avatar tape was significant, $F(8, 78) = 2.616$, $p = .014$, Wilk's $\Lambda = 0.622$, $\eta_p^{2} = .21$, the main effect of emotion was not significant ($p = .137$). 
However, the multivariate main effect of the interaction between emotion and avatar was significant, $F(40, 398) = 2.727$, $p < .001$, Wilk's $\Lambda = 0.616$, $\eta_p^{2} = .22$.

The Greenhouse-Geisser corrected analyses of univariate effects showed that the main effect of the avatar was significant for perception, $p = .035$, $\eta_p^{2} = .25$, but not for elicitation ($p = .227$). 
The effects of emotion were significant for perception ($p = .032$) and for elicitation ($p = .016$). 
The interaction effects were not significant for both perception ($p = .209$) and elicitation ($p = .130$). 
The pairwise comparisons showed that the expression of anger was perceived with significantly more confidence than fear ($p = .006$), and the expression of anger was also perceived with significantly more confidence than disgust ($p = .008$). The corresponding violin plots are also shown in Figure \ref{tab: Violin plots emotion perceiption}. 
Furthermore, we observed a few other significant differences in emotion perception between avatar types. 
The raccoon avatar was perceived significantly more confidently than the banana avatar ($p=.013$), the pig avatar ($p=.011$), and the shark avatar ($p=.028$).
However, differences between the human avatar and the other exemplars did not reach statistical significance (all $p > .082$). 
These results suggest that the type of avatar influenced emotion perception and that certain types of avatars, such as the raccoon, may provide more easily recognizable emotional expressions.

\subsubsection{Emotion elicitation}
\begin{figure*}[htb]
\centerline{\includegraphics[width=1\textwidth]{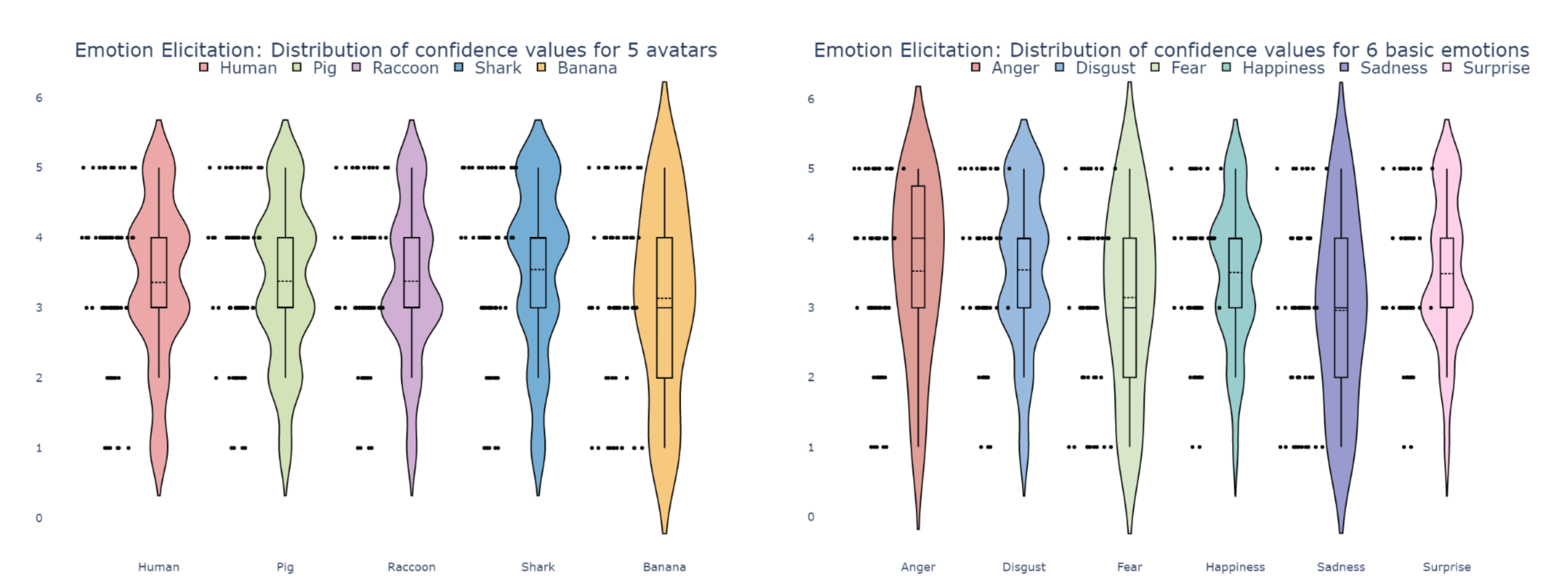}}
\caption{The results of confidence values from 11 participants in task "Emotion elicitation" are shown in violin plots. Left: The results are analysed based on 5 different avatars. From left to right: human: red, pig: green, raccoon: purple, shark: blue, banana: yellow. Right: The results are analysed  based on 6 basix emotions. From left to right: anger: red, disgust: blue, fear: light green, happiness: green, purple: sadness, pink: surprise.}
\label{tab: Violin plots emotion Elicitation}
\end{figure*}

Figure \ref{tab: Violin plots emotion Elicitation} details 11 participants' annotation results in the emotion elicitation task.
As suggested by the results displayed in the right panel of Figure \ref{tab: Violin plots emotion Elicitation}, certain emotions such as "Happiness", "Anger", and "Surprise" appear to more reliably elicit the intended emotions in viewers than "Disgust", "Fear", or "Sadness". 
Furthermore, viewing the pig avatar appeared to elicit more sadness in participants than any other emotion, irrespective of the intended emotional expression displayed by the avatar. 
Finally, the shark and raccoon avatars appeared to elicit more happiness than the other avatars.

For emotion elicitation, a significant pairwise difference was observed between the banana avatar and the human avatar, such that the human avatar elicited more confidence in how the avatar made participants feel ($p=.010$). 
These results suggest that the choice of avatar may have a substantial impact on the emotions elicited in viewers. 
Interestingly, in some cases, this effect might be stronger than that of the emotional expression shown by the avatar.

Conversely, the emotion perception consistency confidence for sharks and pigs was lower than that for bananas, despite both avatars having higher human similarity and facial feature dimensions than bananas. 
This indicates that, as concluded from the initial experiments, lower values of human similarity introduce more ambiguity into emotion perception. This phenomenon only occurred when lower human similarity was the sole factor considered. 

At the end of the experiment, 5 participants responded that they would like to use sharks, 6 participants noted that they found the shark to be the funniest avatar, and 2 participants responded that they liked the shark, and that the shark made them feel happy, suggesting that this could lead to increased tolerance of shark emotions, with the cute avatars being more likely to evoke positive emotions. 
As illustrated in Figure \ref{tab: Emotion Elicitation} below, both sharks and raccoons successfully elicited positive emotions from participants in Task~2 (Emotion elicitation), attributed to their cute characteristics. 
In contrast, the human and pig avatars had the opposite effect. 
Participants observed that the pig avatar, with its natural facial expression conveying sadness, was more likely to evoke negative emotions in Task~2. This observation is further supported by the confusion matrix of emotion elicitation, as depicted in Figure \ref{tab: Emotion Elicitation}. 
It illustrates how people's emotions tend to feel happier when exposed to the cute shark incarnation, despite the shark conveying negative emotions. Conversely, the pig avatar tends to induce more sadness in users, as its natural state is designed to convey sadness.



\begin{figure*}[htb]
\centerline{\includegraphics[width=1\textwidth]{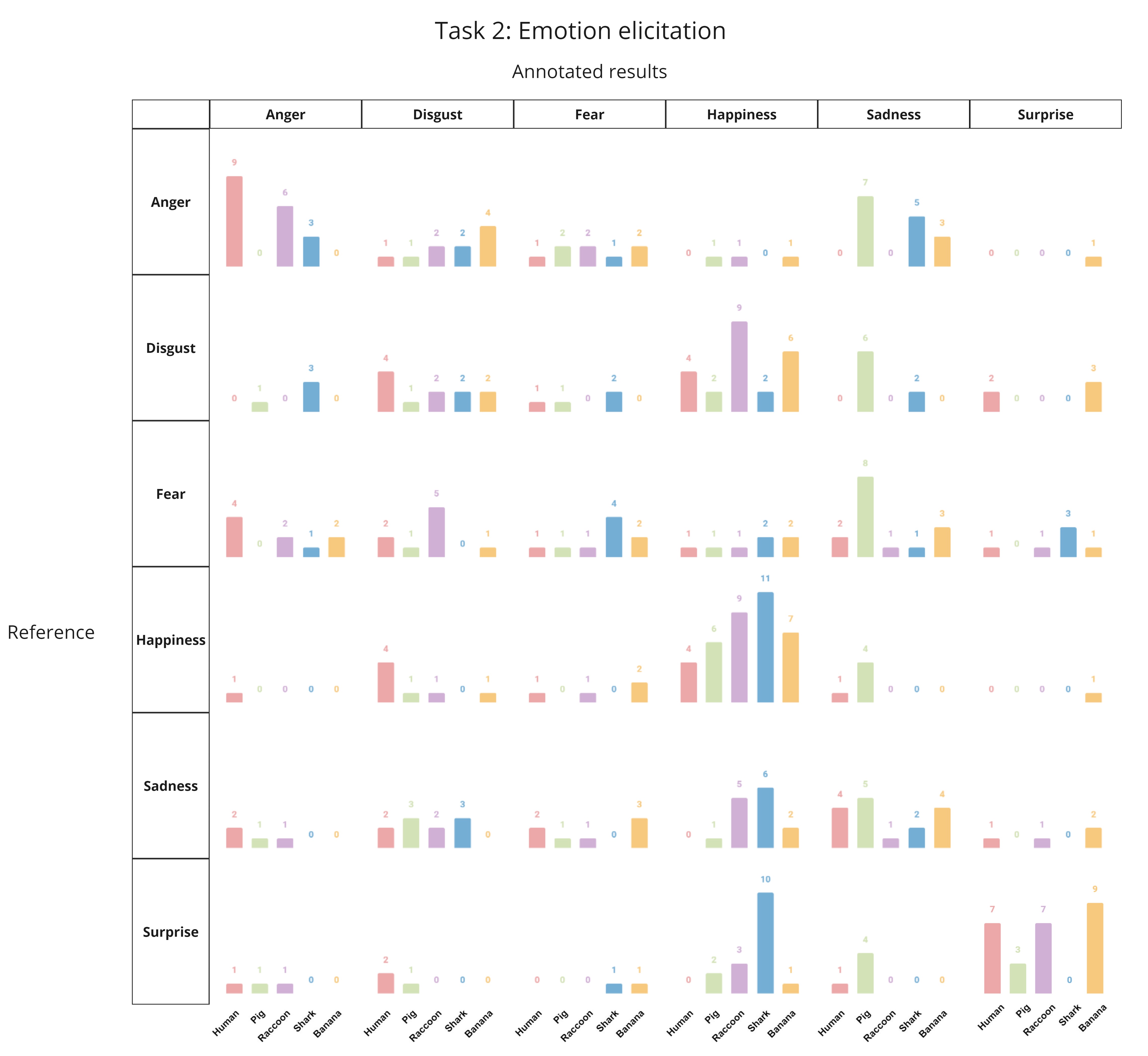}}
\caption{Results of emotion elicitation rated by 11 participants.}
\label{tab: Emotion Elicitation}
\end{figure*}

\section{Discussion}
\label{sec: discussion}
The findings highlight an intriguing and nuanced relationship between an avatar's degree of human-likeness and its ability to convey emotions effectively. 
The results support an intuitive relationship: human avatars have the most human-like characteristics, followed by raccoons and bananas. 
The banana avatar's inability to discriminate between head and trunk suggests a higher level of abstraction, resulting in a more dramatic divergence from human likeness. 
This is further evidenced by the absence of typical human features such as noses and ears. 
Interestingly, despite being the most abstract and least human-like avatar, raters were ambivalent about its ability to convey emotion. 
In contrast, the raccoon avatar, with a lower human likeness score than the human avatar, appeared to communicate the intended emotions most reliably and with relatively few confusions. 
This shows that a certain degree of similarity to human characteristics can help with emotion perception. However, it may not be the only requirement for effective emotional communication.
For instance, as noted in the follow-up extended study, the inherent cuteness of the avatar influences the communication of emotions towards a positive level. 
If the avatar has a naturally sad expression, users can readily perceive it, resulting in a more negative emotional response. 
Interestingly, the dimensions of facial features, which align more closely with humans, did not appear to differ between most of the avatars. However, the reduced visibility of facial features for the banana might still have had an effect on emotion perception and elicitation.

Based on these findings and additional analyses showing low consistency between raters (Fleiss' kappa), our results suggest that emotions expressed by the banana avatar were overall more difficult to recognize. 
However, while participants had difficulty understanding the banana avatar's emotions, they nevertheless expressed relatively high confidence in their emotion perception, suggesting a difference between perceived and factual knowledge. Furthermore, the results suggest a sophisticated interaction between positive and negative emotions. Raccoon avatars appeared to shift raters' perceptions of negative to positive emotions, but human avatars showed the opposite trend. Overall, these results appear to be in line with the concept of Uncanny Valley in aesthetics. 
The concept suggests that humanoid objects that imperfectly resemble actual human beings provoke uncanny or strangely familiar feelings of uneasiness and revulsion in observers. 
"Valley" denotes a dip in the human observer's affinity for the replica, a relation that otherwise increases with the replica's human likeness \cite{mori2012uncanny}. 
Human avatars are too similar to human beings, but there are always differences, so they bring the emotional response of the raters into the valley. 
They may elicit overall more negative emotional responses towards human avatars compared to more abstract types of avatars.

Over 70\% of the participants favored abstract avatars for everyday scenarios, with "cuteness" identified as a key attribute. 
Notably, at the questionnaire's end, participants expressed a preference for the final video to feature a shark avatar, indicating satisfaction with this choice. 
Interestingly, 16.7\% found the pig avatar amusing, yet none wished to use it in daily life. 
Exploring the aspect of prior semiotic stances towards non-human representations in future studies would be worthwhile.

Additionally, during our usage of the software, we noticed a section of the user experience that could be enhanced. 
Despite the software offering numerous avatars, the human avatar features lean towards Western facial characteristics. 
Users with smaller eyes are consistently identified as blinking when facial features are recognized, presenting a potential disruption in the software's usability.

\section{Conclusion and future work}
\label{sec: conclusion}
In summary, while the degree of human-likeness in avatars is important for emotion perception, the results show that it is not the only determining factor. 
Other fundamental characteristics of avatars, such as their perceived nature or innate traits, may be equally important. 
Our findings can help choose the appropriate avatar for different use scenarios, such as live streaming, games, psychological intervention, commercial promotion, and other human-computer interaction scenarios. 
A suitable avatar may affect the user's emotional and psychological state, and proper induction can achieve the purpose of interaction more efficiently.

To better understand how avatars may be used to convey and intervene in users' emotions more accurately, future work should aim to more systematically manipulate dimensions of human-likeness as well as other notable differences between avatars, such as facial width and the size and visibility of features involved in producing facial expressions. 
Here, e.g., the differences observed in the present work between the raccoon and the banana could serve as a starting point. Intriguingly, unlike real-world creatures, virtual avatars may also rely less on conveying emotions with their eyes and most virtual avatars do not have exact eye representation. This could have a substantial influence on the way people are used to recognizing emotions through their eyes. 
Finally, in our final questions at the end of the experiment, and as suggested by further comments from participants after the experiment, we observed that virtual avatars appeared to draw more attention to the changes in mouth and eyebrow regions, in particular during the emotion perception task. We therefore believe that further exploring the role and priority of different facial features in emotional communication may help to further improve virtual avatars and the quality of their facial expressions, so that users can get a more realistic experience when communicating with avatars.

\section*{Acknowledgment}
The research reported on in this paper received funding from the EU’s H2020 RIA programme under grant agreement no. 951846 (MUHAI) and from the Collaborative Research Center (SFB) 1320 EASE – Everyday Activity Science and Engineering, University of Bremen (\url{www.ease-crc.org}), sub-project P01 "Embodied Semantics for the Language of Action and Change".

\bibliographystyle{IEEEtran}
\bibliography{references}
\end{document}